\documentclass[aps,ams,amsmath,prx,longbibliography,twocolumn,superscriptaddress]{revtex4-2}
\usepackage{graphics}
\usepackage{graphicx}
\usepackage{epstopdf,epsfig}
\usepackage{amsmath}
\usepackage{amssymb}
\usepackage{amsfonts}
\usepackage{mathrsfs}
\usepackage{bm}
\usepackage{xcolor}
\usepackage{bbm}
\usepackage{hyperref}
\usepackage[normalem]{ulem}
\usepackage{comment}
\usepackage{soul}
\usepackage[varg]{txfonts}

\definecolor{jm}{rgb}{0.6,0.0,0.0}
\setul{0.5ex}{0.3ex}
\setulcolor{jm}
\setstcolor{jm}

\begin{document}

\title{Electrical magnetochiral anisotropy in Rashba superconductors}

\author{Joaquim Telles de Miranda}
\affiliation{Centro Brasileiro de Pesquisas F\'isicas, Rua Xavier Sigaud 150, 22290-180, Rio de Janeiro, Brazil}
\affiliation{Department of Physics, University of Wisconsin-Madison, Madison, Wisconsin 53706, USA}

\author{Maxim Khodas}
\affiliation{Racah Institute of Physics, Hebrew University of Jerusalem, Jerusalem 91904, Israel}
\affiliation{Materials Science Division, Argonne National Laboratory, Lemont, Illinois 60439, USA}

\author{Alex Levchenko}
\affiliation{Department of Physics, University of Wisconsin-Madison, Madison, Wisconsin 53706, USA}

\date{June 17, 2026}

\begin{abstract}
We theoretically investigate the role of higher-order Lifshitz invariants in nonreciprocal charge transport in two-dimensional noncentrosymmetric superconductors with Rashba spin-orbit coupling. In the superconducting state, these symmetry-allowed terms give rise to critical-current nonreciprocity, while in the normal state near the superconducting transition they generate a pronounced magnetochiral anisotropy. Using symmetry-constrained group-theoretical methods, we systematically construct the allowed Lifshitz invariants and derive the corresponding vector structure of the nonreciprocal current.
To describe nonlinear transport in the fluctuation regime, we apply a generalized time-dependent Ginzburg-Landau theory that incorporates both the Aslamazov-Larkin contribution from fluctuation-induced Cooper pairs and the Maki-Thompson contribution associated with quantum interference in the Cooper channel. We further analyze the effects of disorder scattering and dephasing on the resulting nonreciprocal response.
\end{abstract}

\maketitle

\section{Introduction}

In noncentrosymmetric systems where inversion symmetry ($\mathcal{P}$) is broken, the coupling of a charge carrier's momentum to its spin via an asymmetric crystalline potential gradient generates Rashba spin-orbit coupling (SOC) \cite{Rashba:59,Bychkov:84}. This coupling creates a spin-momentum-locked Fermi surface featuring a characteristic helical spin texture. When an external magnetic field ($\bm{B}$) is applied in the plane of a two-dimensional Rashba conductor, time-reversal symmetry ($\mathcal{T}$) is further broken. The shifting of the spin-locked bands by the in-plane Zeeman energy induces an asymmetric deformation of the spin texture in the momentum space. This results in electrical magnetochiral anisotropy (MCA), where the resistance of the material depends nonreciprocally on the mutual orientation (parallel or antiparallel) of the magnetic field and current \cite{Rikken:01,Rikken:05}.

In the superconducting state, the interplay of the Rashba spin-splitting and the in-plane Zeeman field induces an asymmetric, finite-momentum Cooper pairing state, a realization of the helical Fulde-Ferrell-Larkin-Ovchinnikov (FFLO) state \cite{FF:64,LO:64}. Because the system pairs at an absolute momentum shift $\bm{q}_0 \propto [\bm{B} \times \hat{\bm{z}}]$, the absolute thermodynamic critical current (e.g. depairing maximal current) 
will generally differ for forward  ($j_{c+}$) and backward ($j_{c-}$) flow directions. This manifests the superconducting diode effect (SDE) \cite{Nadeem:23,Ma:25,Shaffer:25}, which can be viewed as a particular example of magnetoelectric effects in superconductors. 

At first sight, SDE and MCA appear to have little in common beyond the fact that both require the breaking of $\mathcal{P}$ and $\mathcal{T}$ symmetries and therefore exhibit nonreciprocal behavior. Indeed, SDE is an equilibrium thermodynamic property of a superconductor, whereas MCA is a dissipative nonlinear transport effect that emerges in the resistive state. The deeper possible connection between these seemingly distinct phenomena, including extension to a broader class of photogalvanic effects, was suggested only relatively recently \cite{Wakatsuki:18,Wakatsuki:18,Hoshino:18,Buzdin:24,Daido:24,JTM:26}. It is rooted in the higher-order Lifshitz invariants, symmetry-allowed contributions to the superconducting free energy that are odd in spatial gradients (or, equivalently, in the collective Cooper-pair momentum $\bm{q}$ of the condensate when presented in Fourier space).

The linear Lifshitz invariant, $\hat{\bm{z}}\cdot[\bm{B}\times\bm{q}]$, is precisely the term that gives rise to the helical state. In the original work by Edelstein, Ref. \cite{Edelstein:96}, it was argued that this term is also responsible for critical-current nonreciprocity. It was later shown in Ref. \cite{Agterberg:12} that this is not the case, the supercurrent remains reciprocal, a consequence of gauge invariance. In an infinite, strictly two-dimensional system, a finite $\bm{q}_0$ is therefore largely inconsequential. The notable exceptions are: (i) a Josephson junction in which the helical state extends over a finite region of length $L$ becomes anomalous, acquiring a phase shift $\phi_0$; the current-phase relation and the Fraunhofer pattern are then governed by the product $q_0L$ \cite{Sigrist:05}; and (ii) the Meissner effect becomes unconventional, with the magnetic field acquiring an oscillatory component as it decays into the bulk of the superconductor \cite{LNE:85}. Consequently, critical-current nonreciprocity requires a systematic treatment of terms cubic in $\bm{q}$ \cite{Yuan:22,Ilic:22,Daido:22,He:22,Hasan:24}.

In nonsuperconducting systems, MCA is typically a small effect in weakly spin-orbit-coupled metals, since the relevant inversion- and time-reversal-breaking energy scales (spin-orbit splitting $\Delta_{\text{SO}}$ and Zeeman energy $\Delta_{\text{Z}}$) are often small compared to the Fermi energy $E_{\text{F}}$. Consequently, nonlinear nonreciprocal coefficients are parametrically suppressed in powers of these small ratios $\{\Delta_{\text{SO}},\Delta_{\text{Z}}\}/E_{\text{F}}\ll1$. For this reason, the observed giant MCA in the superconducting transition metal dichalcogenide MoS$_2$ has attracted significant interest and motivated further studies of nonreciprocal charge transport in superconducting systems; see Refs. \cite{Tokura:18,Nagaosa:24} for recent reviews.

The enhancement of the MCA signal by several orders of magnitude was observed in the normal state in proximity to the superconducting transition near the critical temperature $T_c$ of this material. As is well known, near $T_c$ the transport properties are strongly affected by thermally driven fluctuations of the order parameter, leading to excess conductivity, the so-called paraconductivity \cite{LarkinVarlamov:05}. In linear response theory, it consists of the Aslamazov-Larkin contribution \cite{AL:68}, describing direct transport by fluctuation-induced Cooper pairs, and the Maki-Thompson contribution \cite{Maki:68,Thompson:70}, which accounts for quantum interference processes in the Cooper channel.

Since the typical energy scale of fluctuation-induced pairs is set by the inverse Ginzburg-Landau time, $\Delta_{\text{GL}}=\tau^{-1}_{\mathrm{GL}} \sim T - T_c$, which is much smaller than the Fermi energy, the effects of symmetry-breaking perturbations in the Cooper channel of charge transport are strongly enhanced. It can further be shown that the resulting nonreciprocity is related to the asymmetric momentum dispersion of Cooper pairs, a consequence of the higher-order Lifshitz invariants.

Doped MoS$_2$ is characterized by a basal-plane mirror symmetry $\sigma_h$. In particular, monolayers with prismatic coordination possess the point-group symmetry of a trigonal prism, $D_{3h}$. The $\sigma_h$ symmetry forbids Lifshitz invariants for the in-plane magnetic field. As a consequence, the leading allowed Lifshitz invariant is cubic in momentum, has the structure of a warping term  and requires an out-of-plane magnetic field. 

SDE deep in the superconducting state of MoS$_2$ was not reported in Ref. \cite{Wakatsuki:18}, so it is not possible to establish an unambiguous direct link to the observed MCA as being induced by a particular cubic Lifshitz invariant in the free energy (as opposed to the other possible mechanisms such as parity mixing \cite{Wakatsuki:18,Hoshino:18}). Conversely, SDE has been measured in other superconducting TMDs such as NbSe$_2$ \cite{Bauriedl:22} and in NbS$_2$/NbSe$_2$ heterostructures \cite{Guan:26}, where both out-of-plane and in-plane magnetic fields can independently generate and control SDE. However, MCA has not been independently measured in the fluctuation regime in these systems. Thus, to the best of our knowledge, there is no experiment reporting concurrent measurements of SDE and MCA in the same material. Nevertheless, results from independent studies on Rashba and Ising superconductors, as well as theoretical considerations, including MCA in superconducting surface states of three-dimensional topological insulators such as Bi$_2$Se$_3$ and Bi$_2$Te$_3$, suggest that SDE and MCA likely share a common physical origin.

Motivated by these insights, we undertake a further theoretical study of MCA with a particular focus on Rashba superconductors. Compared to several recent related works \cite{Wakatsuki:18,Hoshino:18,Daido:24}, our analysis serves a number of complementary purposes. First, we employ a symmetry-constrained group-theoretical approach to systematically construct cubic Lifshitz invariants for point groups consistent with broken basal mirror-plane symmetry, for example $C_{3v}$ corresponding to regular triangular pyramid, as well as the continuous symmetry group $C_{\infty v}$ of a cone. The same symmetries also give rise to generalized Rashba spin-orbit splitting of the electronic bands. Next, we apply a generalized time-dependent Ginzburg-Landau (TDGL) framework to describe nonlinear transport in the regime of strong superconducting fluctuations. Our formulation incorporates both the Aslamazov-Larkin and Maki-Thompson contributions to the nonreciprocal current. Using the derived Lifshitz invariants, we then establish the corresponding vector structure of the MCA response and independently verify its form based on symmetry considerations alone. Finally, we perform explicit calculations in the diffusive regime to emphasize the importance of impurity scattering and analyze the sensitivity of the interference contribution to the dephasing time.

\section{TDGL theory of the nonlinear response}

The framework of TDGL theory developed by Gor'kov and Eliashberg \cite{GE:68}, despite its well-known limitations, provides a practical tool for calculating dynamical response functions of superconductors. In particular, it is well suited for describing the fluctuation regime above $T_c$ and superconducting states with sufficiently strong pair breaking. Working in a gauge characterized by a time-dependent vector potential $\bm{A}(t)$, Schmid  \cite{Schmid:69} derived a general expression for the Aslamazov-Larkin fluctuation current by solving the dynamical equation for the superconducting order parameter. The resulting expression takes the form
\begin{align}\label{eq:j-AL}
\bm{j}_{\text{AL}}&=4eT\int_{\bm{q}}\partial_{\bm{q}}\alpha(\bm{q})\nonumber \\ 
&\times\int^{0}_{-\infty}dt_1
\exp\left[-2\int^{0}_{t_1}dt_2\alpha(\bm{q}-2e\bm{A}(t_2+t)+2e\bm{A}(t))\right],
\end{align}
where we introduced a shorthand notation $\int_{\bm{q}}=\int\frac{d^2\bm{q}}{(2\pi)^2}$ to denote the momentum integration.  
The dispersion of the fluctuation-driven Cooper pairs is described by the function $\alpha(\bm{q})$. In the diffusive limit it is given by 
\begin{equation}
\alpha(\bm{q})=Dq^2+\Delta_{\text{GL}}+\delta\alpha(\bm{q})
\end{equation}
where $D$ is the diffusion coefficient and $\Delta_{\text{GL}}=\tau^{-1}_{\text{GL}}=\frac{8}{\pi}(T-T_c)$ is the GL gap. 
For noncentrosymmetric system $\delta\alpha(\bm{q})$ captures Lifshitz invariants and makes the dispersion of fluctuating Cooper pairs to be asymmetric in momentum, $\alpha(\bm{q})\neq\alpha(-\bm{q})$. 

Equation \eqref{eq:j-AL} describes the full nonlinear current. We are interested in its nonreciprocal part, $\delta\bm{j}_{\text{AL}}$, which is quadratic in electric field $\bm{E}$ corresponding to the MCA. To this end, we take $\bm{A}=-\bm{E}t$, and expand Eq. \eqref{eq:j-AL} to the quadratic order in $\bm{E}$. This gives
\begin{equation}
\bm{j}_{\text{AL}}=\sigma_{\text{AL}}\bm{E}+\delta\bm{j}_{\text{AL}}. 
\end{equation}     
The first term describes the AL conductivity \cite{AL:68}
\begin{equation}
\sigma_{\text{AL}}=4e^2TD\int_{\bm{q}}\frac{Dq^2}{(Dq^2+\Delta_{\text{GL}})^3}=\frac{e^2}{2\pi}(T\tau_{\text{GL}}). 
\end{equation} 
The second term reads explicitly 
\begin{align}
\delta\bm{j}_{\text{AL}}=&4eT\int_{\bm{q}}\partial_{\bm{q}}\alpha(\bm{q})\int^{0}_{-\infty}dt_1\exp(2t_1\alpha(\bm{q}))\nonumber\\
&\times\left[\frac{4e^2}{3}E_jE_k\partial_{q_j}\partial_{q_k}\alpha(\bm{q}) t^3_1+2e^2E_jE_k\partial_{q_j}\alpha(\bm{q})\partial_{q_k}\alpha(\bm{q}) t^4_1\right]. 
\end{align}
Here the summation over the repeated index is implicitly assumed. After integration over the time variable it reduces to 
\begin{equation}
\delta\bm{j}_{\text{AL}}=2e^3T\int_{\bm{q}}\partial_{\bm{q}}\alpha(\bm{q})\left[\frac{3\partial_{q_j}\alpha(\bm{q})\partial_{q_k}\alpha(\bm{q})}{\alpha^5(\bm{q})}-\frac{\partial^2_{q_jq_k}\alpha(\bm{q})}{\alpha^4(\bm{q})}\right]E_jE_k
\end{equation}
This expression can be integrated by parts, symmetrized, and finally brought to the form 
\begin{equation}\label{eq:j-AL-MCA}
\delta\bm{j}_{\text{AL}}=e^3T\int_{\bm{q}}\frac{\partial_{\bm{q}}\alpha(\bm{q})\partial_{q_j}\partial_{q_k}\alpha(\bm{q})}{\alpha^4(\bm{q})}E_jE_k
\end{equation}
It can be easily seen now that, if not for the asymmetry term in the dispersion, this expression would vanish, since $\partial_{\bm{q}}\alpha(\bm{q})$ is odd in momentum and thus averages to zero upon momentum integration. To make the result finite, one must retain the additional oddness of $\delta\alpha(\bm{q})$. This means that, for a centrosymmetric system, the expansion starts at cubic order in $\bm{E}$ and therefore describes the nonlinear conductivity, $\bm{j}_{\text{AL}}=\sigma_{\text{AL}}(E)\bm{E}$. Interestingly, the full functional form of the nonlinear paraconductivity $\sigma_{\text{AL}}(E)$ can be obtained analytically, as shown by Dorsey \cite{Dorsey:91}.

The Maki-Thompson term was not originally captured by the TDGL equations developed in Ref.~\cite{GE:68}. For a long time, it was not entirely clear at which stage of the derivation it had been omitted from the formalism. The reason is perhaps subtle. The anomalous MT contribution arises as a correction to the current and possesses a rather intricate causal structure involving both retarded and advanced Cooperon propagators. As a result, it was lost during the analytic continuation procedure. This contribution was subsequently recovered in the calculations of Ref. \cite{Levchenko:07} using the Keldysh formulation of the field theory \cite{Kamenev:09}, where the correct analytic structure is built in from the outset. The component of the current describing quantum interference in the Cooper channel reads  
\begin{align}\label{eq:j-MT}
\bm{j}_{\text{MT}}&=16e^2TD\bm{E}\nonumber \\ 
&\times
\int_{\bm{q}}\int^{t}_{-\infty}dt_1\exp\left[-2\int^{t}_{t_1}dt'\beta(\bm{q}-e\bm{A}(t')-e\bm{A}(2t_1-t'))\right]\nonumber \\ 
&\times \int^{t_1}_{-\infty}dt_2\exp\left[-2\int^{t_1}_{t_2}dt''\alpha(\bm{q}-2e\bm{A}(t'')) \right]
\end{align}
where 
\begin{equation}
\beta(\bm{q})=Dq^2+\tau^{-1}_\phi
\end{equation}
is related to the two-Cooperon part of the anomalous MT diagram, and $\tau_\phi$ denotes the dephasing time. Note that $\bm{j}_{\text{MT}}$ couples directly to the electric field, whereas $\bm{j}_{\text{AL}}$ couples to the collective Cooper-pair momentum. For this reason, one may expect these terms to have different symmetry properties, which should manifest in the nonlinear response.

To the linear order in $\bm{E}$ one can ignore the vector potential in $\alpha$ and $\beta$ functions. Therefore, the nonreciprocal part of the MT current comes from the expansions of these functions to the linear order in $\bm{E}$. Being interested in the most singular term in $\Delta_{\text{GL}}$ one should expand the pair-propagator part of the expression defined by $\alpha(\bm{q})$. By expanding $\beta(\bm{q})$ one gets an additional term which is less sensitive to $T-T_c$ and decays faster with the pair-breaking time. In an analogy with the AL term we split the current into the linear and nonreciprocal parts
\begin{equation}
\bm{j}_{\text{MT}}=\sigma_{\text{MT}}\bm{E}+\delta\bm{j}_{\text{MT}}.
\end{equation}
Performing the expansion over $\bm{E}$ and integration over time, we recover the MT term   
\begin{equation}
\sigma_{\text{MT}}=\int_{\bm{q}}\frac{4e^2TD}{(Dq^2+\Delta_{\text{GL}})(Dq^2+\tau^{-1}_\phi)}=\frac{e^2}{\pi}(T\tau_{\text{GL}})\ln\frac{\tau_\phi}{\tau_{\text{GL}}},
\end{equation}
and more importantly obtain its nonreciprocal part 
\begin{equation}\label{eq:j-MT-MCA}
\delta\bm{j}_{\text{MT}}=4e^3TD\int_{\bm{q}}\frac{\bm{E}(\bm{E}\cdot\partial_{\bm{q}}\alpha(\bm{q}))}{\beta(\bm{q})\alpha^3(\bm{q})}. 
\end{equation}
Again, this term survives only if $\alpha(\bm{q})$ contains odd-in-$\bm{q}$ contributions. Equations \eqref{eq:j-AL-MCA} and \eqref{eq:j-MT-MCA} constitute the central results of this section. They demonstrate that MCA due to superconducting fluctuations is controlled by the Lifshitz invariants in the dispersion of preformed Cooper pairs, which make the spectrum asymmetric.

\section{MCA from Lifshitz invariants}

All linear Lifshitz invariants arising in noncentrosymmetric superconductors have been tabulated in Ref.~\cite{Agterberg:12}. The explicit form of these terms can be established on the basis of the symmetry properties of the crystallographic point groups. Microscopic calculations performed for a variety of model systems \cite{Mineev:94,Edelstein:96,Mineev:08,Houzet:15} confirm their structure and determine the corresponding coefficients, which can be expressed in terms of the microscopic parameters of a given model.

Nonlinear Lifshitz invariants have been studied much less extensively. Most of what is currently known comes from the context of the superconducting diode effect, where cubic terms have been derived using 
Bogoliubov-de Gennes Hamiltonian \cite{Yuan:22}, semiclassical Eilenberger equations \cite{Ilic:22}, detailed microscopic lattice model \cite{Daido:22}, phenomenological approach \cite{He:22}, and direct diagrammatic techniques \cite{Hasan:24}. Here, we adopt a symmetry-constrained group-theoretical approach to determine the allowed vector forms of these invariants, leaving the calculation of the corresponding coefficients for future work. Without any further constraints or limitations, this approach enables us to establish the vector structure of the MCA current and its temperature dependence in the vicinity of $T_c$.

To this end, consider point group $C_{3v}$ which is relevant for the Rashba-Zeeman system. Let us have a $m_x$ mirror as symmetry operation for definiteness. In this case the three copies of $E$ irreducible representations (irreps) transforming identically are $(q_x,q_y)$ and $(2 q_x q_y, q_x^2 - q_y^2)$, and $(-B_y,B_x)$. So we have $q_x^2 + q_y^2 \in A_1$, and $(2 q_x q_y, q_x^2 - q_y^2) \in E$ for the $(q_x,q_y) \otimes (q_x,q_y)$, and similarly $B_x q_y - B_y q_x \in A_1$, $(B_x q_x - B_y q_y, -B_y q_x -B_x q_y ) \in E$ for the decomposition of  $(q_x,q_y) \otimes (B_x,B_y)$. As a result, we can get possibly two scalars. One is from the two $A_1$s
\begin{equation}\label{eq:alpha-C3v}
    \delta\alpha_{C_{3v}}(\bm{q}) = \alpha_3 (q_x^2 + q_y^2) (q_x B_y - q_y B_x).
\end{equation}
The second possible scalar comes from the product of the two $E$s,
$(B_x q_x - B_y q_y) 2 q_x q_y + ( -B_y q_x -B_x q_y)(q_x^2 - q_y^2) = B_x ( q_x^2 q_y +q_y^3) + B_y(- q_y^2 q_x - q_x^3) = (B_x q_y - B_y q_x) (q_x^2 + q_y^2)$
so no other invariants, just the one we had in Eq. \eqref{eq:alpha-C3v}.  

For the point group $C_{4v}$ the $(q_x,q_y)$ transforms as $(-B_y,B_x)$, therefore there are two independent invariants
\begin{equation}\label{eq:alpha-C4v}
  \delta\alpha_{C_{4v}}(\bm{q}) = \alpha_3 (q_x^2 + q_y^2) (q_x B_y - q_y B_x) + \alpha'_3 q_x q_y (q_x B_x - q_y B_y).
\end{equation}
In the point group $C_{6v}$ we have only one term which coincides with that in Eq. \eqref{eq:alpha-C3v}.

We can use now these expressions for $\delta\alpha(\bm{q})$ in Eqs. \eqref{eq:j-AL-MCA} and \eqref{eq:j-MT-MCA}. 
We consider the $C_{4v}$ case as it contains more vector combinations and thus covers all the possibilities in this example. Inserting Eq. \eqref{eq:alpha-C4v}
into Eq. \eqref{eq:j-AL-MCA}, expanding to the linear order in magnetic field, and performing momentum integration we find as a result for the AL contribution  
\begin{equation}\label{eq:j-AL-MCA-fin}
    \delta \bm{j}_\textrm{AL} = \frac{e^3}{12\pi D T_c} (T_c \tau_\textrm{GL})^2 \left[\alpha_3\bm{F}(\bm{E},\bm{B})+\alpha'_3\bm{G}(\bm{E},\bm{B})\right].
\end{equation}
The vectors appearing in this result have the following forms
\begin{subequations}
\begin{equation}
\bm{F}=\left(\begin{array}{c}
3E_x^2B_y - 2E_xE_yB_x+E_y^2B_y \\ -E_x^2B_x +2E_xE_yB_y - 3E_y^2B_x  
\end{array} \right)
\end{equation}
\begin{equation}
\bm{G}=\left(\begin{array}{c}
2E_xE_yB_x - E_y^2B_y \\ E_x^2B_x -2E_xE_yB_y 
\end{array} \right)
\end{equation}
\end{subequations} 
In both $\bm{F}$ and $\bm{G}$ the upper element of the column vector corresponds to the $x$-component of the current, and lower element to the $y$-component.

For the MT contribution, we find that both terms in the expression for $\delta\alpha(\bm{q})$ generate the same vector for the MCA current. It has the form   
\begin{equation}\label{eq:j-MT-MCA-fin}
\delta \bm{j}_\textrm{MT} = \frac{e^3}{\pi D T_c} (T_c \tau_\textrm{GL})^2 f\left(\frac{\tau_\textrm{GL}}{\tau_\phi}\right)\left(\alpha_3 - \frac{\alpha'_3}{4}\right) \bm{E}[\bm{E} \times \bm{B}]_z
\end{equation}
The dimensionless function is found to be 
\begin{equation}
f(x)=\frac{2x(2+x)\ln x+x(4-5x)+1}{2(x-1)^4}. 
\end{equation}
Using the standard vector identity one can show that the transverse (or Hall-like) magnetochiral MT contribution
can be decomposed into two vectors: the longitudinal magnetochiral term and pure Rashba-type term since the following relation holds
\begin{equation}
\bm{E}[\bm{E} \times \bm{B}]_z=(\bm{E}\cdot\bm{B})[\bm{E}\times\hat{\bm{z}}]+E^2[\hat{\bm{z}}\times\bm{B}]. 
\end{equation}

We see that at weak pair breaking, $\Delta_{\text{GL}}\tau_\phi\gg1$, both AL and MT contributions have the same dependence on $\Delta_{\text{GL}}\propto T-T_c$, however at strong pair-breaking, $\Delta_{\text{GL}}\tau_\phi\ll1$, MT term is parametrically suppressed by a factor $(T\tau_\phi)^2\ll1$ 

\section{MCA from symmetry considerations}

The vector structure of the MCA current in this problem can be established on the basis of symmetry arguments, even without an explicit calculation. We present this analysis both for completeness and as a means of cross-checking the final results. To this end, let us rewrite the relevant combinations in a more transparent form for the $C_{4v}$ point group. The product representation $(E_x,E_y)\otimes(E_x,E_y)$ decomposes into the irreducible representations
$
E_x^2+E_y^2 \in A_1,
E_x^2-E_y^2 \in B_1,
E_xE_y \in B_2.
$
These quantities must then be combined with the vector $(-B_y,B_x)$ to construct the current vector $(j_x,j_y)$. The simplest contribution is the isotropic term
$\bm{E}^2 [\hat{\bm{z}} \times \bm{B}]$.

In general, given a vector $(v_x,v_y)$ the vector combinations to the second order in $E$ are
$C_1=  (E_x^2 + E_y^2) (v_x,v_y); C_2 = (E_y^2 - E_x^2) (v_x,-v_y); C_3 = E_x E_y (v_y,v_x)$.
The independent invariants are obtained via identification $(v_x, v_y) = ( - B_y, B_x) $,
so that we get three allowed combinations, 
$C_1=  (E_x^2 + E_y^2) ( - B_y, B_x) ; C_2 = (E_y^2 - E_x^2) (B_y,B_x); C_3 = E_x E_y (B_x,-B_y)$. 
Therefore the total current should be a linear combination,
\begin{align}
    \bm{j}_{\text{MCA}} = q_1 C_1 + q_2 C_2 + q_3 C_3. 
    \end{align}  
To reproduce the form of $\delta\bm{j}_{\text{AL}}$ in Eq. \eqref{eq:j-AL-MCA-fin} one has to have $q_1 = -2$, $q_2 = -1$ and $q_3=-2$ for the $\bm{F}$-vector and  $q_1 = - q_2 = 1/2$, and $q_3 =2$ for the $\bm{G}$-vector.

\section{Summary and discussion} 

In this work, we presented a theoretical study of nonreciprocal charge transport in the regime of superconducting fluctuations. Specifically, we investigated magnetochiral anisotropy (MCA) in superconductors with Rashba spin-orbit coupling and established a direct connection between the superconducting diode effect and MCA through the presence of cubic Lifshitz invariants. These terms emerge in the spectrum of Cooper pairs in noncentrosymmetric superconductors with broken time-reversal symmetry and provide the fundamental source of nonreciprocal transport. Adopting a symmetry-based approach, we systematically constructed the allowed Lifshitz invariants and incorporated them into an extended time-dependent Ginzburg-Landau framework to derive the nonlinear current response. Our formulation captures the leading fluctuation-induced contributions to the current, including both thermal fluctuations of Cooper pairs and their quantum-interference corrections. As an independent consistency check, we verified the resulting vector structure of the MCA current using a separate symmetry-constrained analysis.

We conclude with several remarks regarding the scope and limitations of our approach. Throughout this work, we have not attempted to evaluate the coefficients of the Lifshitz invariants appearing in Eqs.~\eqref{eq:alpha-C3v} and \eqref{eq:alpha-C4v}, as their determination requires a specific microscopic model. For Rashba superconductors, such coefficients have been calculated in both the clean and disordered limits \cite{Ilic:22,Hasan:24,Hasan:25}. For example, in the diffusive regime and for weak spin--orbit coupling, one finds $\alpha_3\propto \alpha_{\text{R}}D(\Delta_{\text{SO}}\tau)^2/T^3_c$, where $\alpha_{\mathrm{R}}$ is the Rashba velocity characterizing the strength of spin-orbit coupling, $\tau$ is the elastic scattering time on impurities, and $\Delta_{\text{SO}}=\alpha_{\text{R}}p_{\text{F}}$ with $p_{\mathrm{F}}$ denoting the Fermi momentum. Notably, this coefficient is strongly suppressed in the disordered limit relative to the clean case by powers of the small parameter $T_c\tau\ll 1$. We emphasize, however, that extending our results to the clean regime is not simply a matter of replacing one set of coefficients by another. Once the electronic mean free path exceeds the superconducting coherence length, the transport problem becomes intrinsically nonlocal. In that regime, the local expressions for the current employed throughout this work are no longer directly applicable, and a more general theoretical framework is required. In principle, it can be done by extending the results of Ref. \cite{AHL:95} to the nonlinear response, which however appears to be challenging task.   

Another subtle point concerns the role of linear Lifshitz invariants. Within the time-dependent Ginzburg-Landau (TDGL) framework, the superconducting transition is associated with the softening of the fluctuation mode that minimizes the pair propagator $\alpha(\bm{q})$. In conventional superconductors, this minimum occurs at $\bm{q}=0$. In contrast, in helical superconductors the presence of a linear Lifshitz invariant shifts the minimum to a finite wave vector $\bm{q}_0$. Consequently, the dominant fluctuation contribution to transport near $T_c$ originates from modes with momenta in the vicinity of $\bm{q}_0$ rather than around the origin. This feature can be incorporated straightforwardly by expanding the Ginzburg-Landau coefficients in powers of $\delta\bm{q}=\bm{q}-\bm{q}_0$, 
$\alpha(\bm{q})\approx \alpha_0+\alpha^{ij}_{2}\delta q_i\delta q_j+\alpha^{ijk}_3 \delta q_i\delta q_j\delta q_k+\ldots$. For brevity, we have omitted this technical step in the main text. Importantly, it does not alter the structure of the resulting expressions for the nonlinear current, but merely renormalizes the corresponding coefficients. For example, the coefficient $\alpha_3$ appearing in Eq.~\eqref{eq:alpha-C3v}  is replaced by $\alpha_3-2\alpha_1\alpha_4/\alpha_2$, while the momentum variable should be understood as measured relative to the helical wave vector, $\bm{q}\rightarrow \bm{q}-\bm{q}_0$.

The framework developed in this work can be naturally extended to investigate a broader class of nonreciprocal transport phenomena in superconductors, including the nonlinear Hall effect, the photogalvanic effect, and second-harmonic generation. Such extensions are particularly relevant in light of the growing family of superconducting transition-metal dichalcogenides and related moir\'e systems, including WTe$_2$ \cite{Cobden:18,Fatemi:18}, MoTe$_2$ \cite{Kang:15,Qi:16}, and moir\'e-engineered twisted WSe$_2$ \cite{Xia:25}. Several studies along these directions have already been reported \cite{Daido:24,Parafilo:22,Dong:25,Matsumoto:25}, highlighting the richness of nonlinear superconducting transport phenomena in noncentrosymmetric materials. Given the rapid pace of experimental and theoretical developments in this field, we anticipate that many further results will emerge in the near future.

\section*{Acknowledgements}

The work of J. T. M. was supported in part by the Coordenação de Aperfeiçoamento de Pessoal de Nível Superior - Brasil (CAPES) - Finance Code 001 and by the NSF Quantum Leap Challenge Institute for Hybrid Quantum Architectures and Networks Grant No. OMA-2016136. M. K. acknowledges the support of the grant NSF-BSF DMR2023693.
The work of A. L. was supported by NSF Grant No. DMR-2452658 and H. I. Romnes Faculty Fellowship provided by the University of Wisconsin-Madison Office of the Vice Chancellor for Research and Graduate Education with funding from the Wisconsin Alumni Research Foundation. This work was performed in part during the workshop program "Emerging New Phases in Quantum Materials: The Disordered, the Strange and the Topological" at the Aspen Center for Physics, which is supported by National Science Foundation grant PHY-2210452. 


\bibliography{biblio-MCA-Rashba-SC}

\end{document}